# Superluminal pulse reflection from a weakly absorbing dielectric slab

Li-Gang Wang and Shi-Yao Zhu

*Department of Physics, Zhejiang University, Hangzhou, 310027, China*

*and Department of Physics, Hong Kong Baptist University, Kowloon Tong, Hong Kong*

Group delay for a reflected light pulse from a weakly absorbing dielectric slab is theoretically investigated, and large negative group delay is found for weak absorption near a resonance of the slab ( $\text{Re}(kd) = m\pi$ ). The group delays for both the reflected and transmitted pulses will be saturated with the increase of the absorption. © 2006 Optical Society of America.

*OCIS codes*: 320.7120, 310.6870, 260.2110, 230.5750.

The control of pulse propagation in different kinds of media has received extensively attentions for several decades. [1-4] It is now well established that the group delay (also known as the phase time) of the reflected and transmitted pulses can be manipulated from subluminal (slow) to superluminal (fast) propagation by controlling the dispersive properties of a medium. Most literatures have been focused on the transmitted pulse from a medium (including the bulk materials [3-4] and the photonic layer structures [1-2]). Recently, superluminal reflected pulses were also theoretically studied in different circumstances: such as the unstable regime of optical phase-conjugating mirror,[5] low-finesse Fabry-Perot cavities (with silver mirrors) containing absorbing atoms,[6] a dielectric slab system doped with absorptive two-level or three-level atoms,[7] asymmetric photonic band gaps,[8] active fiber Bragg gratings,[9] single quantum wells[10] and asymmetric single quantum barriers.[11] In experiments, Vetter et al.[12] verified the existence of the negative phase time for scattering at quantum wells from a microwave analogy experiment. Longhi et al.[13,14] have also observed superluminal reflected pulse from a double-Lorentzian fiber Bragg grating. Recently, Gevorgyan[15] found the anomalies of radiation absorption and superluminal propagation of light in an isotropic layer. Li[16] noticed that the reflected wave from a lossless dielectric slab undergoes a phase discontinuity (a sudden phase change of $\pi$) at resonant transmission, and he pointed out that the discontinuity of the reflected phase with *null* reflection has no practical meaning. In the view of continuity, we expect that there is a large finite slope of the change of phase with a *nonzero*, albeit small, reflection if the dielectric slab is weakly absorbing. In this Letter, we report that the negative group delay, which can be large, for the reflected pulse near resonances from the weakly absorbing dielectric slab. The negative group delay indicates that the superluminal pulse reflection can be observed when a light pulse is reflected from the weakly absorbing dielectric slab. We also theoretically demonstrate the saturation effect of both the reflected and transmitted group delays with the increase of the absorption.

Consider a light pulse normally incident on the weakly absorbing dielectric slab (extended from $z=0$ to $z=d$ in the $z$ direction) with the complex relative permittivity, $\varepsilon = \varepsilon_r + i\varepsilon_i$, where $\varepsilon_i$ represents the absorption, and $\mu \equiv 1$, as shown in Fig. 1. Both sides of the slab are vacuum. The total electric and magnetic fields from the $z=0$ to $z=d$ can be related via a transfer matrix[17-18]

$$\begin{pmatrix} \cos[kd] & i\frac{1}{q}\sin[kd] \\ iq\sin[kd] & \cos[kd] \end{pmatrix}, \qquad (1)$$

where $k=(\omega/c)\sqrt{\varepsilon}$ is the complex wave number in the slab, $c$ is light speed in the vacuum, $q=\sqrt{\varepsilon}$ for TE polarization, and $q=1/\sqrt{\varepsilon}$ for TM polarization. In the following, we consider the case for TE-polarized plane-wave pulse. Similar results can also be obtained for the TM-polarized plane-wave pulses. Then the reflection and transmission coefficients for TE waves can be evaluated with the help of the transfer matrix method [17, 19]

$$r(\omega,d) = \frac{-(i/2)\left(\frac{1}{\sqrt{\varepsilon}}-\sqrt{\varepsilon}\right)\sin(kd)}{\cos(kd)-(i/2)\left(\frac{1}{\sqrt{\varepsilon}}+\sqrt{\varepsilon}\right)\sin(kd)}, \qquad (2)$$

$$t(\omega,d) = \frac{1}{\cos(kd)-(i/2)\left(\frac{1}{\sqrt{\varepsilon}}+\sqrt{\varepsilon}\right)\sin(kd)}. \qquad (3)$$

We assume that the incident pulse is a Gaussian pulse. In the limit of the very narrow spectrum, $\Delta\omega \ll \omega_c$ (where $\Delta\omega$ is the spectral width of the incident pulse, and $\omega_c$ is the carrier frequency, i. e., both the reflected and transmitted pulses suffer nominal distortion), the group delays for the reflected and transmitted pulse are defined by [6, 20, 21]

$$\tau_{r,t} = [\partial\phi_{r,t}/\partial\omega]_{\omega=\omega_c}, \qquad (4)$$

where $\phi_{r,t}$ are the phases of the transmission and the reflection coefficients, $r(\omega,d)$ and $t(\omega,d)$, respectively, namely, $r(\omega,d) =| r(\omega,d) | \exp[i\phi_r(\omega,d)]$ and $t(\omega,d) =| t(\omega,d) | \exp[i\phi_t(\omega,d)]$.

Define $g_1 \equiv |g_1| e^{i\phi_1} = \cos(kd) - \frac{i}{2}\left[(1/\sqrt{\varepsilon}) + \sqrt{\varepsilon}\right]\sin(kd)$, we have $\tau_t = \partial\phi_t/\partial\omega = -\partial\phi_1/\partial\omega$. With these relations, the transmitted group delay can be analytically expressed as

$$\tau_t = \frac{d}{c}\{4n_i^2(|\varepsilon|-1)\cos[2\operatorname{Re}(kd)] - 4n_r^2(|\varepsilon|+1)\cosh[2\operatorname{Im}(kd)]$$
$$+ n_i[(|\varepsilon|-1)^2 - 4n_i^2]\sin[2\operatorname{Re}(kd)] - n_r[(|\varepsilon|+1)^2 + 4n_r^2]\sinh[2\operatorname{Im}(kd)]\}$$
$$/\{[(|\varepsilon|-1)^2 - 4n_i^2]\cos[2\operatorname{Re}(kd)] - [(|\varepsilon|+1)^2 + 4n_r^2]\cosh[2\operatorname{Im}(kd)]$$
$$- 4n_i(|\varepsilon|-1)\sin[2\operatorname{Re}(kd)] - 4n_r(|\varepsilon|+1)\sinh[2\operatorname{Im}(kd)]\}$$

(5)

here $n_r \equiv \operatorname{Re}(\sqrt{\varepsilon})$ and $n_i \equiv \operatorname{Im}(\sqrt{\varepsilon})$ are, respectively, the real and imaginary parts of the complex refractive index of the slab. Define $g_2 \equiv |g_2| e^{i\phi_2} = -(i/2)[(1/\sqrt{\varepsilon}) - \sqrt{\varepsilon}]\sin(kd)$, from Eq. (2), we have the relation:

$$\tau_r = \tau_t + \tau_1, \tag{6}$$

where $\tau_1$ is analytically given by

$$\tau_1 \equiv \partial\phi_2/\partial\omega = \frac{d}{c}\frac{n_i \sin[2\operatorname{Re}(kd)] - n_r \sinh[2\operatorname{Im}(kd)]}{\cosh[2\operatorname{Im}(kd)] - \cos[2\operatorname{Re}(kd)]}. \tag{7}$$

For a lossless slab system ($\varepsilon_i \equiv 0$) we have $\varepsilon \equiv \varepsilon_r$, $n_i = 0$, and $\sinh[2\operatorname{Im}(kd)] = 0$. In this case, we can find that $\tau_1$ is always equal to zero from Eq. (7), so the reflected group delay is equal to the transmitted group delay. This has been noticed in many other situations [1, 20, 22]. From Eqs.(5-7), we can obtain $\tau_r = \tau_t = 4(d/c)\varepsilon_r(1+\varepsilon_r)/[1 + 6\varepsilon_r + \varepsilon_r^2 - (\varepsilon_r - 1)^2 \cos(2kd)]$. Obviously, for $kd = m\pi$ ($m$ integers), both $\tau_r$ and $\tau_t$ are enhanced to their maximums: $(\tau_{r,t})_{\max} = \frac{1+\varepsilon_r}{2n_r}\tau_0$, where $\tau_0 = d/(c/n_r)$ is the time delay expected from the phase velocity of light traveling through the lossless dielectric slab; for $kd = (m+1/2)\pi$, both $\tau_r$ and $\tau_t$ are suppressed to their minimums: $(\tau_{r,t})_{min} = \frac{2n_r}{1+\varepsilon_r}\tau_0$. However, for the reflected group delay, it has no physical meaning at resonances $kd = m\pi$, owing to the zero reflection [see Fig. 2(b)] and the undefined phases [see the sudden phase changes on the dotted curve of $\phi_r$ in Fig. 2(c)]. There is a large finite slope of the change of phase with a *nonzero*, albeit small, reflection if the slab is weakly absorbing (with complex $\varepsilon$). This large finite slope will lead to the large advancement of $\tau_r$ for the reflected pulse. Actually, in Eq. (6) the first term $\tau_t$ is always positive even for the absorptive dielectric slab. It is the second term $\tau_1$ that leads the reflected group delay to be negative. Especially, at resonances of $\text{Re}(kd) = m\pi$, $\tau_1$ can be simplified to $\tau_1 = -\tau_0 \coth[\text{Im}(kd)]$. It is clear that $\tau_1$ can be greatly negative for $\text{Im}(kd) << 1$.

Figure 2 shows the typical group delay of a reflected pulse from the slab with thickness $d = 6$ $\mu$m and $\varepsilon_r = 3.0$, where $\varepsilon_i = 0$ (dotted), 0.01 (solid), 0.02 (dash-dotted), 0.05 (dashed) in Fig 2a, with the reflection coefficient $|r|$ and the relative phases $\phi_r$ in Fig. 2b and 2c. It is seen that the reflected group delay $\tau_r$ becomes large negative near the resonant frequencies of the slab with the fixed thickness. For the lossless slab ($\varepsilon_i = 0$), we have $|r|=0$ and an abrupt phase jump at the resonance. This effect is similar to that of lateral shift occurring in a lossless slab.[16,19] For the absorbing slabs, the phase change at resonances becomes continuous with a large negative slope, which leads to the large negative group delay near resonances with a nonzero reflection. From Fig. 2b, it is found that at resonances the value of $|r|$ decreases as $\varepsilon_i$ decreases, while the negative group delay of the reflected pulse becomes larger and larger (due to the steeper phase change with respect to angular frequency $\omega$), see Fig. 2a. We believe this giant negative group delay can be observed experimentally. The negative group delay for the reflected pulse could also occur in the case of the inclined incidence, where the resonant condition becomes $\text{Re}(k_z d) = m\pi$ with $k_z$ being the z component of the wave vector in the absorbing dielectric slab. This effect is very similar to that in asymmetric single quantum barrier,[11] and is also analogy to the lateral shift of a light beam reflecting from an absorptive slab. [19]

In Fig. 3(a) we plot the dependence of the group delays on the slab thickness for the reflected and transmitted pulses from a slab with $\varepsilon_r = 3.0$ and $\varepsilon_i = 0.02$ at the frequency $\omega_c / 2\pi = 129.9$ THz. It is clear that, owing to the absorption of the medium, the group delay of the reflected pulse is not equal to that of the transmitted pulses, and the group delay of the reflected pulse becomes large negative when the thickness $d$ approaches the resonance, while the group delay of the transmitted pulse is still positive. In Fig. 3(b) we plot the dependence of the reflected and transmitted group delays on the absorptive parameter $\varepsilon_i$ with $d = 6$ $\mu$m (very close to the resonance of $m = 9$), and other parameters are the same as in Fig. 3(a). It can be seen that near resonance the reflected group delay becomes a very large negative when the absorption is sufficiently weak. For example, for $\varepsilon_i = 0.04$ we have $\tau_r \approx -0.15$ ps, while for $\varepsilon_i = 0.013$ the reflected group delay becomes much more negative, $\tau_r \approx -0.53$ ps. Both the reflected and transmitted group delay will be saturated with the increasing of the absorption. Similar saturation effects in $|r|$ and $|t|$ as the increase of $\varepsilon_i$ are also clear in Fig, 3. This is due to that an increase of the absorption leads to an increase in the effective barrier thickness, resulting in the saturation effect of the reflected and transmitted group delays as in the Hartman effect.[23]

We have theoretically shown the negative reflected group delay of a light pulse reflected from the weakly absorbing dielectric slab near the resonance $\text{Re}(kd) = m\pi$. The amount of the negative group delay of the reflected pulse becomes very large when the absorption of the dielectric slab is sufficiently weak; at the same time, the intensity of the reflected pulse becomes weaker. We also demonstrate the saturation effects of both the reflected and transmitted group delays with the absorbing parameter.

This work was supported by by National Natural Science Foundation of China (No. 10547138) and RGC (HKBU2027/04P) and NSFC05-06/01. Li-Gang Wang's e-mail address is sxwlg@yahoo.com.cn.

**References**

[1] R. Y. Chiao and A. M. Steinberg, in Progress in Optics, E. Wolf, ed. (Elsevier, Amsterdam, 1997), Vol. **37**, pp. 345-405.

[2] G. Nimtz, Prog. Quant. Electro. **27**, 417 (2003).

[3] R. W. Boyd and D. J. Gauthier, in Progress in Optics, E. Wolf, ed. (Elsevier, Amsterdam, 2002), Vol. **43**, pp. 497-530.

[4] P. W. Milloni, J. Phys. B: At. Mol. Phys. **35**, R31 (2002).

[5] M. Blaauboer, A. G. Kofman, A. E. Kozhekin, G. Kurizki, D. Lenstra, and A. Lodder, Phys. Rev. A **57**, 4905 (1998).

[6] V. S. C. Manga Rao, S. Dutta Gupta, and G. S. Agarwal, Opt. Lett. **29**, 307 (2004).

[7] L. G. Wang, H. Chen, and S. Y. Zhu, Phys. Rev. E **70**, 066602 (2004).

[8] S. Longhi, Phys. Rev. E **64**, 037601 (2001).

[9] S. Longhi, Phys. Rev. E **72**, 056614 (2005).

[10] C. F. Li, Q. Wang, Phys. Lett. A **275**, 287 (2000).

[11] C. F. Li and H. Spieker, Opt. Commun. **259**, 158 (2006).

[12] R. M. Vetter, A. Haibel, G. Nimtz, Phys. Rev. E **63**, 04671 (2001).

[13] S. Longhi, M. Marano, P. Laporta, M. Belmonte, and P. Crespi, Phys. Rev. E **65**, 045602(R) (2002).

[14] S. Longhi, M. Marano, M. Belmonte, and P. Laporta, IEEE J. Sel. Top. Quantum Electron. **9**, 4 (2003).

[15] A. H. Gevorgyan, Optics and Spectroscopy, **96**, 877 (2004).

[16] C. F. Li, Phys. Rev. Lett. **91**, 133903 (2003).

[17] L. -G. Wang, N. -H. Liu, Q. Lin and S. -Y. Zhu, Phys. Rev. E **70**, 016601 (2004).

[18] M. Born and E. Wolf, *Principles of optics*, 7th (expanded) ed., (Cambridge University press, Cambridge, 1999).

[19] L. G. Wang, H. Chen, and S. Y. Zhu, Opt. Lett. **30**, 2936 (2005).

[20] A. M. Steinberg and R. Y. Chiao, Phys. Rev. A **49**, 3283 (1994).

[21] E. P. Wigner, Phys. Rev. **98**, 145 (1955).

[22] H. G. Winful, Phys. Rev. Lett. **90**, 055501 (2003).

[23] T. E. Hartman, J. App. Phys. **33**, 3427 (1962).
9

# Figure Captions

FIG. 1. Schematic of the weakly absorbing dielectric slab.

FIG. 2. (Color online) (a) group delay $\tau_r$, (b) absolute value of reflection coefficient, (c) the relatively reflected phase $\phi_r$ as functions of angular frequency $\omega$ with the thickness $d = 6\mu m$ and $\varepsilon_r = 3.0$. Solid curves, $\varepsilon_i = 0.01$; dash-dotted curves, $\varepsilon_i = 0.02$; dashed curves, $\varepsilon_i = 0.05$. Dotted curves for the lossless slab ($\varepsilon_i = 0$).

FIG. 3. (a) Dependence of the reflected (solid curve) and the transmitted (dashed curve) group delay on the slab thickness $d$ with $\varepsilon = 3.0 + 0.02i$ and $\omega_c/2\pi = 129.9\,\text{THz}$. (b) Group delays $\tau_{r,t}$ and (c) absolute values $|r|$, $|t|$ of reflection and transmission coefficients as functions of the absorptive parameter $\varepsilon_i$ at the fixed thickness $d = 6\mu m$. Note the different time scales for $\tau_{r,t}$ in (b). Other parameters are as in Fig. 3 (a).

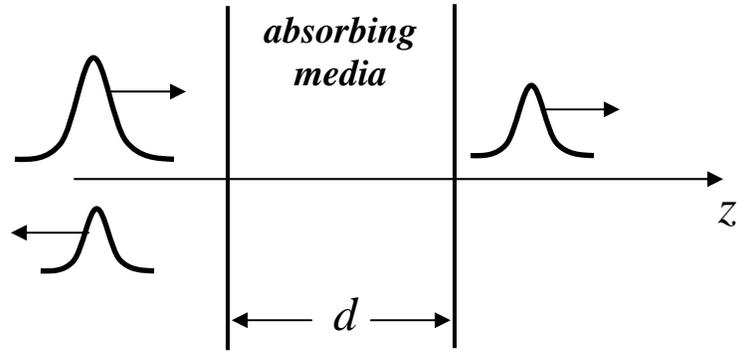

**FIG. 1**

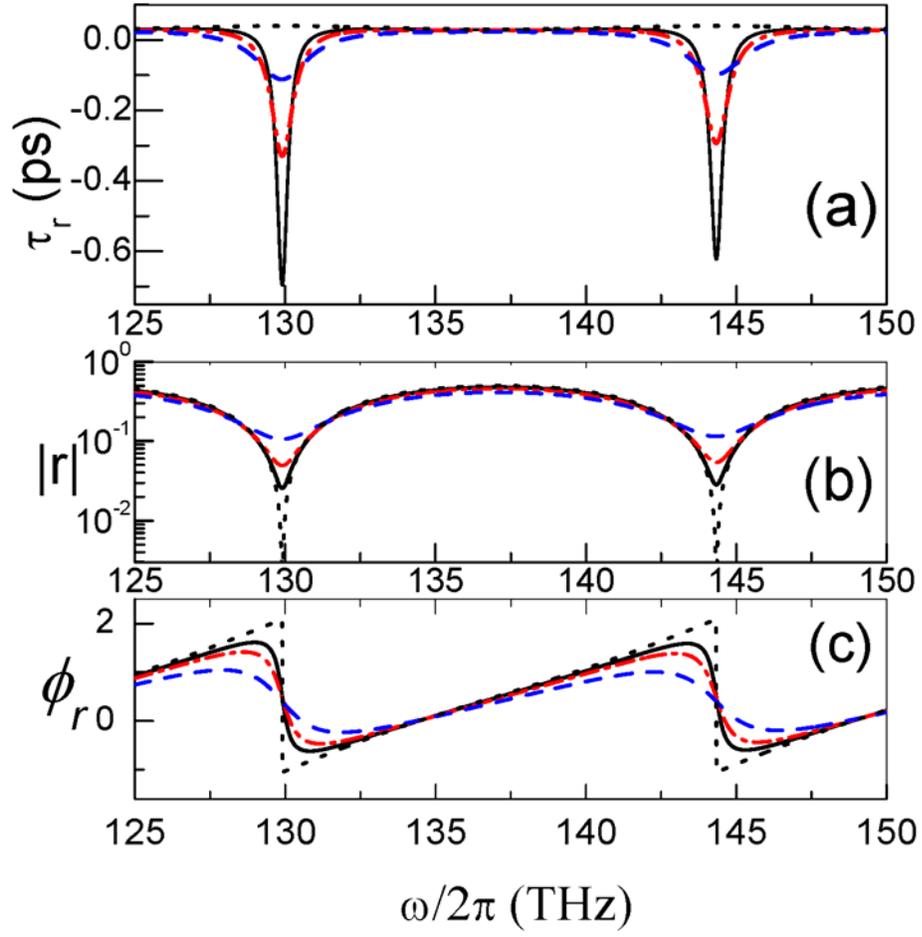

FIG. 3. (Color online)

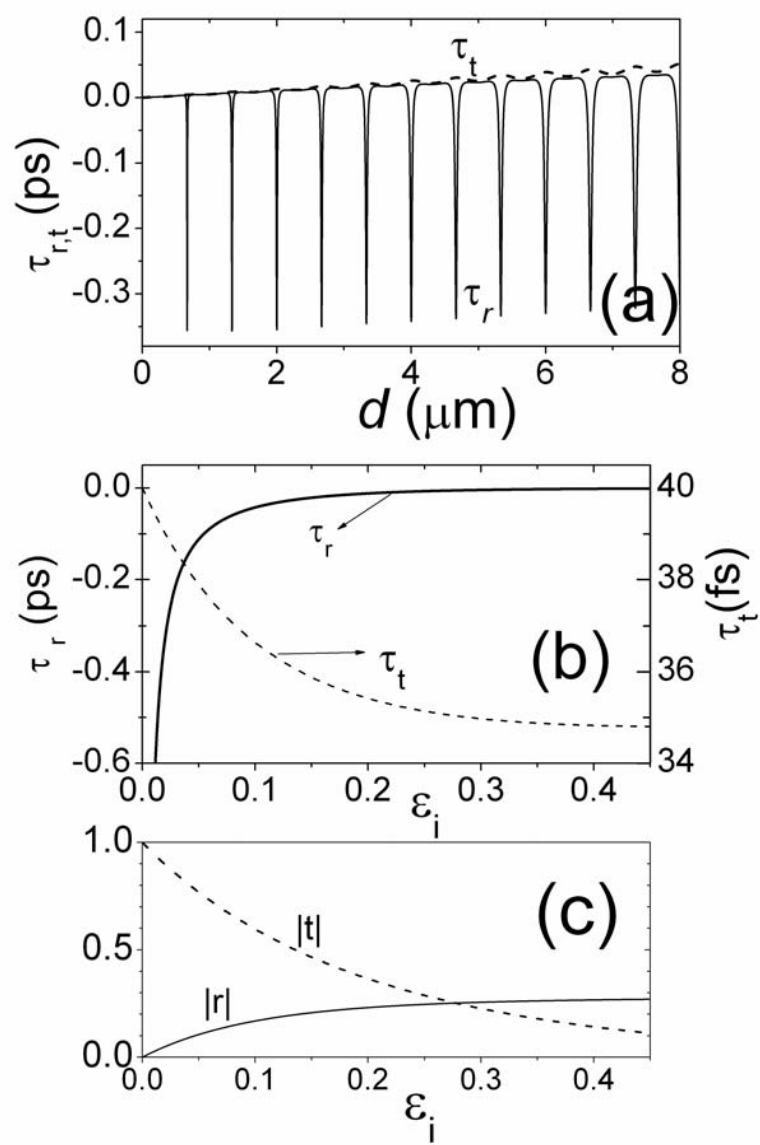

**FIG. 3**